\documentstyle[aps,prl,preprint,epsf,floats,psfig]{revtex}
\tightenlines
\def\square{\vcenter{\vbox{\hrule height.3pt
          \hbox{\vrule width.3pt height6pt
          \kern6pt\vrule width.3pt}\hrule height.3pt}}}

\def\sumint{\hbox{$\sum$}\!\!\!\!\!\!\int}
\newcommand{\beq}{\begin{equation}}
\newcommand{\eeq}{\end{equation}}
\newcommand{\bqa}{\begin{eqnarray}}
\newcommand{\eqa}{\end{eqnarray}}

\begin{document}

\preprint{
\vbox{\halign{&##\hfil\cr
&ITF-UU-01/10\cr
        & hep-ph/yymmnn \cr
&\today\cr }}}

\title{Thermal Effects in Low-Temperature QED}
\author{Jens O. Andersen}
\address{Institute for Theoretical Physics, University of Utrecht,\\
       Leuvenlaan 4, 3584 CE Utrecht, The Netherlands}

\maketitle

\begin{abstract}
{\footnotesize 
QED is studied at low temperature ($T\ll m$, where $m$ is the electron mass) 
and zero chemical potential. 
By integrating out the electron field and the nonzero bosonic Matsubara modes,
we construct an effective three-dimensional field theory that is
valid at distances $R\gg1/T$.
As applications, we reproduce the ring-improved free energy
and calculate the Debye mass to order $e^5$.
}\end{abstract}

\newpage

\section{Introduction}
If we have a quantum field theory in equilibrium at 
temperature $T$, the abundance of particles of mass $m$ much larger than
$T$ is Boltzmann suppressed. More surprisingly, perhaps, is the fact that
there are additional effects that are suppressed only by powers of 
$T/m$~\cite{barton,finn,bj}.
The Boltzmann-suppressed terms can be associated with loop
integrations that involve distribution functions of the heavy particle.
On the other hand, the power-suppressed terms arise from perturbative
corrections involving only distribution functions of light particles, with 
masses on the order of $T$ or less.

Effective field theory ideas suggest that one integrates out the heavy
particle of mass $m$ to construct a low-energy effective field theory
that can be used for $T\ll m$~\cite{lepage,kaplan}. 
Such an approach was used by Kong and 
Ravndal~\cite{finn} to study QED at low temperature
(see also Refs.~[6-13] and Refs. therein
for various calculations in low-temperature QED). 
By integrating out
the electron field, they constructed a low-energy effective field theory
for photons. Since this procedure was carried out at zero temperature,
dimensional analysis tells us that the coefficients in the effective theory
are suppressed by inverse powers of the electron mass $m$.
These higher order interactions are induced by the coupling of the photon
to virtual electron-positron pairs in the vacuum. 
Since the momenta of the photons are on the order $T$ and thus
much smaller than $m$, the electron-positron pairs are off their mass shell
by an amount $\sim m$. Thus they can only propagate a distance $R\sim1/m$
and their effects can be mimicked by local interactions.
Using this effective field theory to calculate corrections to the 
Stefan-Boltzmann law for the pressure, 
they showed that the leading correction goes like $\alpha^2T^8/m^4$.
This correction was obtained by a straightforward
two-loop calculation in the effective theory.
In full QED, it would require a three-loop calculation.

If we are interested only in power corrections, we can determine the
coefficients in the effective field theory by matching at zero temperature.
In the case of QED, this would lead to the Euler-Heisenberg 
Lagrangian~\cite{uh,eh}
with additional higher order operators that can be written in terms of the
field strength and its dual~\cite{finn} (see e.g.~\cite{dicus}
for such higher order operators). 
However, it can be shown that this effective 
Lagrangian leads to a vanishing Debye mass to all orders in perturbation 
theory. We know that this is incorrect, but 
one can account for Debye screening and other Boltzmann-suppressed
effects in low-temperature QED by carrying out the
matching at finite temperature. Alternatively, one may reorganize the
usual perturbative series in QED using resummed propagators in the usual
way. However, such an approach is normally more cumbersome in
practical calculations than effective field theory.

If we are interested in static quantities such as the pressure or Debye
mass, it proves useful to construct a second effective field theory 
for the zero Matsubara mode and this is done by integrating out the nonzero 
Matsubara modes\cite{landsman,gins,BN,kaja}.
This effective
field theory is three-dimensional
and is valid at distances $R\gg 1/T$.
It can be
constructed in a two-step process by first integrating out the
electron field, and then integrating out the nonzero bosonic Matsubara modes.
From a calculational point of view, however, it is easier to integrate
out the fermions and the nonzero Matsubara modes at the same time. 
In this paper, we will take the latter approach.

The paper is organized as follows. 
In section II, we discuss
QED at low temperature
and the construction of the three-dimensional effective field
theory.
In section III, we determine the coefficients in the effective field theory.
In section IV, we apply the effective field theory to calculate the
the free energy to order $e^3$ and the Debye mass to order $e^5$.
Finally, in section V, we summarize and
conclude. All necessary details are collected in three appendices.

\section{QED at Low Temperature}
In the imaginary-time formalism, one can view a quantum field theory in four 
dimensions as a field theory in three Euclidean dimensions with an 
infinite tower of fields, where the Matsubara frequencies act as masses
in the propagators~\cite{landsman}. 
In low-temperature QED, this implies that the fermions
have masses of order $m$, while the nonzero bosonic modes have masses
of order $T$. The zero-frequency bosonic modes are massless.
Thus for distances $R\gg1/T$,
we can construct an effective three-dimensional field theory
for the zero Matsubara modes by integrating out the electron field
as well as the 
nonzero bosonic modes~\cite{landsman,gins,BN,kaja}.
The coefficients of this effective field theory then encode the physics 
at the momentum scales $m$ and $T$.
This procedure is briefly explained in section III.

The partition function of QED can be written as a path integral
\bqa
{\cal Z}=\int{\cal D}\bar{\eta}\;{\cal D}\eta\;
{\cal D}A_{\mu}\;{\cal D}\bar{\psi}\;{\cal D}\psi\;
\exp{\left[-\int_0^{\beta}d\tau\int d^3x\;{\cal L}\right]}\;,
\eqa
where the Euclidean Lagrangian is
\bqa
{\cal L}_E=\frac{1}{4}F_{\mu\nu}F_{\mu\nu}+m\overline{\psi}\psi
+\overline{\psi}
\gamma_{\mu}\Big (\partial_{\mu}-ieA_{\mu} \Big )\psi
+\left(\partial_{\mu}\bar{\eta}\right)\left(\partial_{\mu}\eta\right)
+{\cal L}_{\mbox{\footnotesize gf}}\;.
\eqa
Here, ${\cal L}_{\mbox{\footnotesize gf}}$ denotes the gauge-fixing
term. In the following we work in Feynman gauge, where
\bqa
{\cal L}_{\mbox{\footnotesize gf}}=
\frac{1}{2}(\partial_{\mu}A_{\mu})^{2}\;.
\eqa
However, we emphasize that physical quantities 
are independent of the gauge-fixing condition.

In the three-dimensional effective theory, we can write the partition function
as
\bqa
\label{efff}
{\cal Z}=e^{-f(\Lambda)V}\int{\cal D}\bar{\eta}\;{\cal D}\eta\;
{\cal D}\bar{A}_0\;{\cal D}\bar{A}_{i}\;
\exp{\left[-\int d^3x\;{\cal L}^{}_{\rm eff}\right]}\;,
\eqa
where the prefactor $f(\Lambda)$ is interpreted as the 
coefficient of the unit operator in the effective three-dimensional field 
theory. It depends on an ultraviolet cutoff $\Lambda$ that cancels
the cutoff-dependence in the path integral in Eq.~(\ref{efff})~\cite{BN}.
${\cal L}_{\rm eff}$ is the Lagrangian of the 
effective three-dimensional 
field theory. The effective field theory 
consists of a gauge field $\bar{A}_i$ coupled to a real massive
self-interacting scalar field $\bar{A}_0$
\footnote{The fact that we must allow for terms such 
as~${1\over2}M^2(\Lambda)\bar{A}_0^2$ 
and ${\lambda_3(\Lambda)\over24}\bar{A}_0^4$, is a direct consequence
of the breakdown of Lorentz invariance at finite temperature.}. 
These fields 
can up to normalizations
be identified with the zero-frequency modes of the gauge field in QED.
We can then schematically write
\bqa
\label{eff2}
{\cal L}_{\rm eff}^{}&=&\frac{1}{4}F_{ij}F_{ij}+{1\over4}a_3(\Lambda)
F_{ij}\nabla^2F_{ij}+
{1\over2}(\partial_i\bar{A}_0)^2+{1\over2}M^2(\Lambda)\bar{A}_0^2
+{\lambda_3(\Lambda)\over24}\bar{A}_0^4+
\left(\partial_i\bar{\eta}\right)\left(\partial_i\eta\right) \\ \nonumber
&&
+{\cal L}_{\rm gf}+\delta{\cal L}^{}_{\rm eff}\;,
\eqa
where $\delta{\cal L}^{}_{\rm eff}$ represents all 
higher order local terms that can be 
constructed out of $\bar{A}_i$ and $\bar{A}_0$ 
and respect the symmetries of the theory.
Examples of such symmetries are three-dimensional
gauge invariance and rotational symmetry. This includes renormalizable
terms such as $F_{ij}^2$ as well as nonrenormalizable terms such as
$\bar{A}_0^6$. 
Note also that we have suppressed the $\Lambda$-dependence of the fields
$\bar{A}_i$ and $\bar{A}_0$ in Eq.~(\ref{eff2}).
Let us finally look at the power-counting rules for the effective theory
Eq.~(\ref{eff2})
The coefficients of the operators are power series in $e^2$ 
since we are ignoring infrared divergences and are using conventional
perturbation theory to determine them (see Sec.~\ref{sh}).
Physical quantities are expressed in powers of
the parameters $f(\Lambda)$, $a_3(\Lambda)$, $M(\Lambda)$,... and we must 
figure out
at which order the operators that multiply them start to contribute.
Each power of momentum in a loop integral in the effective theory
gives a factor of $M$, in particular the measure gives a factor $M^3$.
If we want to calculate the free energy to order $e^3$, it is necessary
to determine $f(\Lambda)$ to order $e^2$. Moreover, the one-loop
contribution to the free energy in the effective theory is proportional
to $M^3$, and we therefore need to know the mass parameter to
order $e$. 
(Note that the operator $F_{ij}\nabla^2F_{ij}$
does not contribute to the free energy
at order $e^2$ since it involves only massless fields
whose loop integral vanishes in dimensional regularization. In fact, this
operator can be transformed away by a field redefinition at the expense
of modifying the coeffcients of higher order operators.).
On the other hand, the operator $\bar{A}_0^4$ starts to
contribute first at order
$e^6$; its coefficient $\lambda_3(\Lambda)$ goes like $e^4$ and 
it gives rise to a two-loop diagram in the effective theory where each
loop is proportional to $M$. 
In this manner we can determine at which order in $e$ an operator starts to
contribute to a given physical quantity, and in Eq.~(\ref{eff2})
we have explicitly displayed those operators needed to determine the
free energy to order $e^3$ and the Debye mass to order $e^5$.

\section{Short-distance Coefficients}\label{sh}
In this section, we determine the short-distance coefficients in the 
effective Lagrangian Eq.~(\ref{eff2}). These coefficients must be tuned
as functions of $e$, $T$, and the ultraviolet cutoff $\Lambda$ so
that the effective theory reproduces correlation functions at distances much 
larger than $1/T$. We can carry out these 
calculations using conventional perturbation theory, which is an expansion
in powers of $e^2$. This expansion is plagued with infrared 
divergences due to long-range forces mediated by the massless photon.
These divergences are Debye screened, but can only be taken into account
by resummation. Although the naive perturbative expansion breaks down
due to these infrared divergences,
it can still be used to determine the short-distance coefficients~\cite{BN}.
As long as we treat the long-distance physics in the same incorrect way
using the effective theory, the infrared divergences will cancel in the
matching equations and the coefficients properly encode the short-distance
physics~\cite{BN}.
The Lagrangian of QED is split the usual way 
into free and interacting pieces 
\bqa
{\cal L}_E^{\rm free}&=&\frac{1}{4}F_{\mu\nu}F_{\mu\nu}
+\overline{\psi}\left(m+
/\!\!\!\partial 
\right)\psi
+\left(\partial_{\mu}\bar{\eta}\right)\left(\partial_{\mu}\eta\right)
+{\cal L}_{\mbox{\footnotesize gf}}\;,
\\
{\cal L}_E^{\rm int}&=&-ie/\!\!\!\!A 
\overline{\psi}\psi\;,
\eqa
while the Lagrangian Eq.~(\ref{eff2}) is split according to 
\bqa
{\cal L}_{\rm eff}^{\rm free}&=&
{1\over4}F_{ij}F_{ij}+{1\over2}(\partial_i\bar{A}_0)^2
+\left(\partial_{\mu}\bar{\eta}\right)\left(\partial_{\mu}\eta\right)
+{\cal L}_{\rm gf}
\;,
\\
{\cal L}_{\rm eff}^{\rm int}&=&
{1\over4}a_3(\Lambda)
F_{ij}\nabla^2F_{ij}
+{1\over2}M^2(\Lambda)
\bar{A}_0^2+{\lambda_3(\Lambda)\over24}
\bar{A}_0^4
+\delta{\cal L}_{\rm eff}^{}\;.
\eqa
\subsection{Coefficient of the Unit Operator}
\begin{figure}[htb]
\begin{center}
\mbox{\psfig{figure=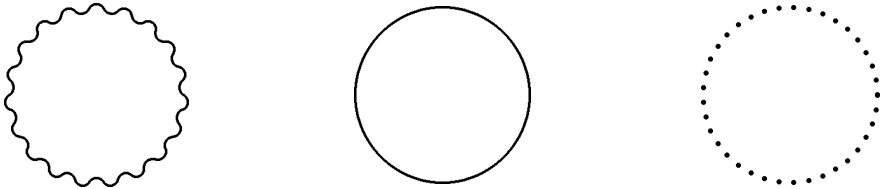}}
\end{center}
\caption[One-loop vacuum diagrams in QED.]{\protect One-loop vacuum diagrams in QED.}
\label{q1}
\end{figure}
\begin{figure}[htb]
\begin{center}
\mbox{\psfig{figure=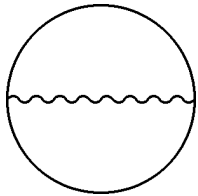}}
\end{center}
\caption[Two-loop vacuum diagram in QED.]{\protect Two-loop vacuum diagram  in QED.}
\label{q2}
\end{figure}
The matching condition that determines the coefficient of the unit 
operator is~\cite{BN}
\bqa
\log{\cal Z}=-f(\Lambda)V+\log{\cal Z}_{\rm eff}^{}\;.
\eqa
The contributions to $\log{\cal Z}$ through order $e^2$ is given
by the Feynman graphs in Figs.~\ref{q1} and~\ref{q2}.
A wavy line denotes a photon, a solid line denotes a fermion, 
and a dotted line denotes a ghosts.
The expression is
\bqa\nonumber
{T\log{\cal Z}\over V}&\approx&-{1\over2}(d-1)\sumint_{P}\log P^2+
2\sumint_{\{P\}}\log(P^2+m^2)
\\
&&
\label{two}
+
{1\over2}e^2\sumint_{\{PQ\}}\mbox{Tr}
\left[\gamma_{\mu}{P\!\!\!\!/-m\over P^2+m^2}
\gamma_{\mu}{Q\!\!\!\!/-m\over Q^2+m^2}
{1\over (P+Q)^2}
\right]\;,
\eqa
where the trace is over Dirac indices.
The sign $\approx$ is a reminder that an equality only holds
in strict perturbation theory.
The fermionic one-loop diagram has a pole in $\epsilon$, which is 
proportional to $m^4$. This pole is cancelled
by the one-loop vacuum counterterm $\Delta_1{\cal E}_0$. 
The two-loop diagram also has 
poles in $\epsilon$. The temperature-independent pole is cancelled
by the two-loop vacuum counterterm $\Delta_2{\cal E}_0$, while the
temperature-dependent ones are cancelled by the one-loop counterterm
diagrams. The two-loop sum-integrals is briefly discussed in appendix A. 
Our renormalization prescription is that the renormalized
vacuum energy vanishes at 
the scale $\Lambda$. Thus the fermionic one-loop diagram and the two-loop
diagram are given by their finite-temperature pieces
(See Eqs.~(\ref{fermip}) and~(\ref{fermi2}) in appendix A)
After renormalization,
the resulting expression in the low-temperature limit is
\bqa
\label{free1}
{T\log{\cal Z}\over V}&\approx&{\pi^2T^4\over 45}
+{4\over(2\pi)^{3/2}}m^{3/2}T^{5/2}e^{-m/T}
+{e^2\over2(2\pi)^3}m^{2}T^2e^{-2m/T}
\;.
\eqa
In the effective theory, all loop diagrams vanish with dimensional
regularization, since there is no mass scale in the 
corresponding integrals
Hence, $\log{\cal Z}_{\rm eff}$ vanishes identically in strict perturbation
theory.
The matching equation then implies 
\bqa
f(\Lambda)=-{\log{\cal Z}\over V}
\;,
\eqa
where the right hand side is given by Eq.~(\ref{free1}).
\subsection{Field Normalization Constant}
\begin{figure}[htb]
\begin{center}
\mbox{\psfig{figure=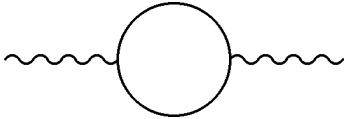}}
\end{center}
\caption[One-loop polarization tensor in QED.]{\protect One-loop polarization tensor in QED.}
\label{pollen}
\end{figure}
%
The field normalization constants for $A_i$ and $A_0$ 
are obtained by reading off the coefficients of
$\delta_{ij}k^2-k_ik_j$ and $k^2$ 
of $\Pi_{ij}^{(1)}(0,{\bf k})$ and $\Pi_{00}^{(1)}(0,{\bf k})$,
where $\Pi_{\mu\nu}^{(1)}(k_0,{\bf k})$ is the 
one-loop polarization tensor~(\ref{pol}).
These are given in Eqs.~(\ref{space}) and~(\ref{stat}) 
in appendix C, and we obtain
\bqa
\label{i}
\bar{A}_i(\Lambda)&\approx&{1\over\sqrt{T}}
\left[1+{4\over3}e^2\sumint_{\{P\}}{1\over(P^2+m^2)^2}\right]^{1/2}A_i
\;,\\ 
\label{0}
\bar{A}_0(\Lambda)&\approx&{1\over\sqrt{T}}
\left[1+{1\over3}e^2\sumint_{\{P\}}
{6\over(P^2+m^2)^2}-{8p_0^2\over(P^2+m^2)^3}\right]^{1/2}A_0\;.
\eqa
Eqs.~(\ref{i}) and~(\ref{0}) have poles in $\epsilon$ that are
cancelled by the wave function renormalization
counterterm 
\bqa
Z_{A}=1-{e^2\over12\pi^2\epsilon}\;.
\eqa
After renormalization, we obtain
\bqa
\label{redef2}
\bar{A}_i(\Lambda)
&\approx&{1\over\sqrt{T}}\left[1+{e^2\over24\pi^2}\left(L-J_2\right)\right]A_i\;, \\
\bar{A}_0(\Lambda)&\approx&{1\over\sqrt{T}}\left[1+{e^2\over24\pi^2}
\left(
L-J_2-J_3m^2T^{-2}
\right)
\right]A_0\;.
\eqa
where $L=\log\left({\Lambda^2\over m^2}\right)$ and the integrals
$J_n$ are defined in appendix A.
Note the different normalizations of the fields $A_0$ and $A_i$. 
In the low-temperature limit, this reduces to
\bqa
\bar{A}_i(\Lambda)&\longrightarrow&{1\over\sqrt{T}}\left[1+{e^2\over24\pi^2}
\left(L-2(2\pi)^{1/2}m^{-1/2}T^{1/2}e^{-m/T}\right)\right]A_i\;, \\
\bar{A}_0(\Lambda)&\longrightarrow&{1\over\sqrt{T}}\left[1+{e^2\over24\pi^2}
\left(
L-(2\pi)^{1/2}m^{1/2}T^{-1/2}e^{-m/T}
\right)
\right]A_0\;.
\eqa

\subsection{Mass Parameter}
The mass parameter can be determined in several ways. We determine it by
demanding that the screening mass in strict perturbation theory be the
same in QED and the effective theory. 
The screening mass $m_s$ is defined as the location of the pole in the 
propagator
for spacelike momentum~\cite{BN}:
\bqa
\label{scr}
k^{2}+\Pi_{00} (0,{\bf k})=0,\hspace{1cm}k^{2}=-m^{2}_{s}\;.
\eqa In the effective theory, we have
\bqa
k^{2}+M^{2}(\Lambda)+\Pi_{\rm eff}(k,\Lambda )=0,
\hspace{1cm}k^{2}=-m^{2}_{s}\;,
\eqa
where $\Pi_{\mbox{\footnotesize eff}}(k,\Lambda )$
is the self-energy of $\bar{A}_0$ in the effective theory. 
We can expand the self-energy function $\Pi_{00}(0,{\bf k})$ in powers
of the external momentum ${\bf k}$. To second order in the loop expansion, 
the solution to Eq.~(\ref{scr}) for the screening mass is~\cite{Andersen}
\bqa
m_s^2\approx\left[1-\Pi_{00}^{(1)\prime}(0,0)
\right]\Pi_{00}^{(1)}(0,0)+\Pi_{00}^{(2)}(0,0)
\;.
\eqa
Here $\Pi_{00}^{(n)}$ denotes the nth order contribution to the static
polarization tensor in the loop expansion,
and the prime denotes differentiation with respect to $k^2$.
The self-energy function $\Pi_{\mbox{\footnotesize eff}}(k,\Lambda )$
can also be expanded in a Taylor series around ${\bf k}=0$. The 
corresponding loop integrals are evaluated at ${\bf k}=0$ and since there
is no other mass scale, the self-energy function $\Pi_{\rm eff}(0,\Lambda)$
vanishes identically
in strict
perturbation theory. Thus the matching condition reduces to
\bqa
M^2(\Lambda)\approx m_s^2\;.
\eqa 
The one-loop contribution to $\Pi_{00}(0,0)$ is given by 
by the first term in Eq~(\ref{stat}), while
$\Pi_{00}^{\prime}(0,0)$ is given by 
the second term in Eq.~(\ref{stat}).
The two-loop contribution to $\Pi_{00}(0,0)$ is given by Eq.~(\ref{twopol}),
and in appendix B, we explain how one can obtain the expression for it from
the two-loop contribution to the free energy.
In the low-temperature limit, the mass parameter becomes 
\bqa
M^2(\Lambda)={4e^2\over(2\pi)^{3/2}}m^{3/2}T^{1/2}e^{-m/T}
-{4e^4\over3(2\pi)^{7/2}}m^{3/2}T^{1/2}Le^{-m/T}
+{10e^4\over3(2\pi)^3}m^2e^{-2m/T}
\;.
\eqa
Using the renormalization group equation for the running gauge coupling,
\bqa
\label{rg}
\mu{de^2\over d\mu}={e^4\over6\pi^2}\;,
\eqa
we see that the mass parameter
is independent of the renormalization scale 
$\mu$ to order $e^4$.

\subsection{Coupling Constants}
\begin{figure}[htb]
\begin{center}
\mbox{\psfig{figure=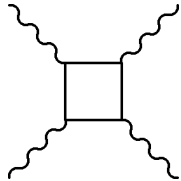}}
\end{center}
\caption[Four-point function with external timelike photons at one-loop in QED.]{Four-point function with external timelike photons at one-loop in QED.}
\label{2lqed}
\end{figure}

The one-loop Feynman diagram for the four-point
function $\Gamma^{(4)}$ with external
time-like photons is shown in
Fig~\ref{2lqed}. The matching equation is simply~\cite{Andersen}
\bqa
\lambda_3(\Lambda)=T\Gamma^{(4)}(0,0,0,0)\;,
\eqa
where the arguments indicate that the loop diagram is to be evaluated at
zero external momenta
The expression for the diagram is
\bqa\nonumber
\Gamma^{(4)}(0,0,0,0)&=&
6e^2\sumint_{\{P\}}\mbox{Tr}
\Bigg[
\gamma_0{P\!\!\!\!/-m\over P^2+m^2}
\gamma_0{P\!\!\!\!/-m\over P^2+m^2}
\gamma_0{P\!\!\!\!/-m\over P^2+m^2}
\gamma_0{P\!\!\!\!/-m\over P^2+m^2}
\Bigg]\\
\label{g4}
&=&
24e^2\sumint_{\{P\}}
\Bigg[
m^4+2m^2P^2-8m^2p_0^2+P^4-8p_0^2P^2+8p_0^4
\Bigg]{1\over(P^2+m^2)^4}\;.
\eqa
The particular combination of sum-integrals in Eq.~(\ref{g4}) 
is finite with dimensional
regularization and we obtain
\bqa
\label{lambda}
\lambda_3(\Lambda)=
-{8e^4\over(2\pi)^2}
J_4m^4T^{-3}\;.
\eqa
The fact that this coefficient vanishes identically
at zero temperature, simply
reflects the gauge invariance of QED.
In the low-temperature limit, Eq.~(\ref{lambda}) reduces to
\bqa
\lambda_3(\Lambda)=-{4e^4\over(2\pi)^{3/2}}m^{3/2}T^{-1/2}
e^{-m/T}\;.
\eqa

The coupling constant $a_3$ is not affected by the field redefinition
Eqs.~(\ref{redef2}) to leading order in $e$, 
so its value is given directly by the
coefficient of the last term in Eq.~(\ref{space}) 
\bqa
a_3(\Lambda)&=&{8\over15}{e^2}\sumint_{\{P\}}
{1\over(P^2+m^2)^3}
\\
\label{a}
&=&{e^2\over60\pi^2m^2}\left[1-m^2T^{-2}J_3\right]\;.
\eqa
In the low-temperature limit, this reduces to
\bqa
a_3(\Lambda)={e^2\over60\pi^2m^2}\left[1-(2\pi)^{1/2}m^{1/2}T^{-1/2}
e^{-m/T}\right]\;.
\eqa
The first term is the standard zero-temperature Uehling term, while 
the second is a new thermal correction.

\section{Free Energy and Debye Mass}
In this section, we use the effective Lagrangian Eq.~(\ref{eff2})
to calculate the free energy to order $e^3$ 
and the Debye mass to order $e^5$, respectively.
In order to take electric screening properly into account, we must 
include the effects of the mass term $M$ to all orders in perturbation theory.
Thus the Lagrangian is split accordingly:
\bqa
{\cal L}_{\rm eff}^{\rm free}&=&
{1\over4}F_{ij}F_{ij}+{1\over2}(\partial_i\bar{A}_0)^2+{1\over2}M^2(\Lambda)
\bar{A}_0^2
+\left(\partial_{\mu}\bar{\eta}\right)\left(\partial_{\mu}\eta\right)
+{\cal L}_{\rm gf}
\;,
\\
{\cal L}_{\rm eff}^{\rm int}&=&{1\over4}a_3(\Lambda)
F_{ij}\nabla^2F_{ij}
+{\lambda_3(\Lambda)\over24}\bar{A}_0^4
+\delta{\cal L}_{\rm eff}^{}\;.
\eqa
The $e^3$ contribution to the free energy is given by a simple 
one-loop calculation in the effective theory: 
\bqa
\label{free3}
{T\log{\cal Z}_{\rm eff}\over V}=
-{1\over2}T\int_{p}\log(p^2+M^2)
-{1\over2}(d-2)T\int_{p}\log p^2
+{1\over2}a_3(\Lambda)(d-2)\int_p p^2
\;.
\eqa
The total free energy is then given by
\bqa
{\cal F}=f(\Lambda)T-{T\log{\cal Z}_{\rm eff}\over V}\;,
\eqa
where $f(\Lambda)$ is given by Eq.(\ref{free1}).
Using Eq.~(\ref{3d}) in appendix B and the expression for the 
mass parameter $M(\Lambda)$ to leading order, Eq.~(\ref{free3}) reduces to
\bqa
\label{free33}
{T\log{\cal Z}_{\rm eff}\over V}=
{4e^3m^{9/4}T^{7/4}\over3(2\pi)^{13/4}}e^{-3m/2T}
\;.
\eqa
The total free energy density is minus the sum of 
Eqs.~(\ref{free1}) and~(\ref{free33}):
\bqa
\label{ftotal}
{\cal F}=-{\pi^2T^4\over45}
-{4\over(2\pi)^{3/2}}m^{3/2}T^{5/2}
e^{-m/T}
-{e^2\over2(2\pi)^3}m^2T^2
e^{-2m/T}
-{4e^3m^{9/4}T^{7/4}\over3(2\pi)^{13/4}}e^{-3m/2T}\;.
\eqa
Eqs.~(\ref{ftotal}) is in agreement with the result first
obtained by Gell-Mann and Bru\"ckner~\cite{gell} in nonrelativistic QED.
Note the term that is non-analytic in $e^2$. It arises from the
summation of an infinite number of infrared divergent loops
(ring diagrams).

The Debye mass $m_{\rm D}$ is given by the location of the pole 
in the propagator
\bqa
\label{pole}
k^2+M^2(\Lambda)
+\Pi_{\rm eff}({\bf k},\Lambda)=0\;,\hspace{1cm}k=im_{\rm D}\;,
\eqa
where $\Pi_{\rm eff}({\bf k},\Lambda)$
denotes the self-energy function of $\bar{A}_0$.
To leading order in $e$, the solution to Eq.~(\ref{pole})
is simply $m_{\rm D}^2=M^2(\Lambda)$. 
The one-loop approximation to the self-energy is
\bqa
\Pi^{(1)}_{\rm eff}({\bf k},\Lambda)
={1\over2}\lambda_3(\Lambda)\int_{\bf p}{1\over p^2+M^2}\;.
\eqa
$\Pi^{(1)}_{\rm eff}({\bf k},\Lambda)$
is independent of the external momentum, so the 
solution to Eq.~(\ref{pole}) is simply
\bqa
m_{\rm D}^2=M^2(\Lambda)
+\Pi^{(1)}_{\rm eff}({\bf k},\Lambda)\;.
\eqa
Using Eq.~(\ref{bi3}) and expanding the mass parameter
$M(\Lambda)$ in powers of $e$, we obtain 
the Debye mass through order $e^5$:
\bqa\nonumber
m_{\rm D}^2
&=&
{4e^2\over(2\pi)^{3/2}}m^{3/2}T^{1/2}e^{-m/T}
-{4e^4\over3(2\pi)^{7/2}}m^{3/2}T^{1/2}Le^{-m/T}
+{10e^4\over3(2\pi)^3}m^2e^{-2m/T}
\\&&
\label{debyef}
+{2e^5\over(2\pi)^{13/4}}m^{9/4}T^{-1/4}e^{-3m/2T}
\;.
\eqa
Eq.~(\ref{debyef}) is the main result of the present paper.
Using the renormalization group equation (\ref{rg}) 
for the running gauge coupling,
we see that the Debye mass is independent of the renormalization scale 
$\Lambda$ up to corrections of order $e^6$.
Note also that there is no term proportional to $e^3$ in the expression
for the Debye mass.
The reason is that there are no bosonic propagators
in the one-loop self-energy graph in QED and fermions
need no resummation, since their Matsubara frequencies
are never zero. 
\section{Summary}
In the present work, we have studied QED at low temperature.
We have constructed an effective field theory in three dimensions
that is valid at distances $R\gg1/T$ by integrating
out the electron field and the nonzero Matsubara modes.
The three-dimensional field theory was used to calculate
the pressure to order $e^3$ and the Debye mass to order $e^5$. 
The pressure and the Debye mass
can be calculated either by resummation or by effective 
field theory. Not only does effective field theory simplify
these calculations, but it
also unravels the contributions to physical quantities 
from different momentum scales.

\section*{Acknowledgments}
This work was supported by the Stichting voor
Fundamenteel Onderzoek der Materie
(FOM), which is supported by the Nederlandse Organisatie voor Wetenschapplijk
Onderzoek (NWO). 

\appendix
\section{Sum-integrals}
\renewcommand{\theequation}{\thesection.\arabic{equation}}
\setcounter{equation}{0}

In the imaginary-time formalism for thermal field theory, 
the 4-momentum $P=(p_0,{\bf p})$ is Euclidean with $P^2=p_0^2+{\bf p}^2$. 
The Euclidean energy $p_0$ has discrete values:
$p_0=2n\pi T$ for bosons and $p_0=(2n+1)\pi T$ for fermions,
where $n$ is an integer. 
Loop diagrams involve sums over $p_0$ and integrals over ${\bf p}$. 
With dimensional regularization, the integral is generalized
to $d = 3-2 \epsilon$ spatial dimensions.
We define the dimensionally regularized sum-integral by
\bqa
  \hbox{$\sum$}\!\!\!\!\!\!\int_{P}& \;\equiv\; &
  \left(\frac{e^\gamma\mu^2}{4\pi}\right)^\epsilon\;
  T\sum_{p_0=2n\pi T}\:\int {d^{3-2\epsilon}p \over (2 \pi)^{3-2\epsilon}}\;,\\ 
  \hbox{$\sum$}\!\!\!\!\!\!\int_{\{P\}}& \;\equiv\; &
  \left(\frac{e^\gamma\mu^2}{4\pi}\right)^\epsilon\;
  T\sum_{p_0=(2n+1)\pi T}\:\int {d^{3-2\epsilon}p \over (2 \pi)^{3-2\epsilon}}\;,
\label{sumint-def}
\eqa
where 
$\mu$ is an arbitrary
momentum scale. 
The factor $(e^\gamma/4\pi)^\epsilon$
is introduced so that, after minimal subtraction 
of the poles in $\epsilon$
due to ultraviolet divergences, $\mu$ coincides 
with the renormalization
scale of the $\overline{\rm MS}$ renormalization scheme.
\subsection{One-loop Sum-integrals}
The specific one-loop 
fermionic
sum-integrals needed are
\bqa
\label{fermip}
\sumint_{\{P\}}
\log(P^2+m^2)&=&{1\over(4\pi)^2}
\left({\mu\over m}\right)^{2\epsilon}
\left[-
{e^{\gamma\epsilon}\Gamma(1+\epsilon)\over\epsilon(1-\epsilon)(2-\epsilon)}
m^4+J_0T^4\right]
\;,\\
\label{spat1}
\sumint_{\{P\}}{1\over P^2+m^2}&=&{1\over(4\pi)^2}
\left({\mu\over m}\right)^{2\epsilon}
\left[-
{e^{\gamma\epsilon}\Gamma(1+\epsilon)\over\epsilon(1-\epsilon)}
m^2-J_1T^2\right]
\;,\\
\sumint_{\{P\}}{1\over(P^2+m^2)^2}&=&{1\over(4\pi)^2}
\left({\mu\over m}\right)^{2\epsilon}
\left[
{e^{\gamma\epsilon}\Gamma(1+\epsilon)\over\epsilon}
-J_2\right]\;,\\
\sumint_{\{P\}}{1\over(P^2+m^2)^3}&=&{1\over2(4\pi)^2}
\left({\mu\over m}\right)^{2\epsilon}
\left[
{e^{\gamma\epsilon}\Gamma(1+\epsilon)}
m^{-2}-J_3T^{-2}\right]
\;,\\
\sumint_{\{P\}}{1\over(P^2+m^2)^4}&=&{1\over6(4\pi)^2}
\left({\mu\over m}\right)^{2\epsilon}
\left[
{e^{\gamma\epsilon}\Gamma(1+\epsilon)}\left(1+\epsilon\right)
m^{-4}-J_4T^{-4}\right]
\;,\\
\label{spat2}
\sumint_{\{P\}}{p_ip_j\over(P^2+m^2)^2}&=&{\delta_{ij}\over2(4\pi)^2}
\left({\mu\over m}\right)^{2\epsilon}
\left[-
{e^{\gamma\epsilon}\Gamma(1+\epsilon)\over\epsilon(1-\epsilon)}m^2
-J_1T^2
\right]
\;,\\
\sumint_{\{P\}}{p_0^2\over(P^2+m^2)^2}&=&{1\over2(4\pi)^2}
\left({\mu\over m}\right)^{2\epsilon}
\left[-
{e^{\gamma\epsilon}\Gamma(1+\epsilon)\over\epsilon(1-\epsilon)}m^2
+(d-2)J_1T^2+2J_2m^2
\right]
\;,\\
\sumint_{\{P\}}{p_0^2\over(P^2+m^2)^3}&=&{1\over4(4\pi)^2}
\left({\mu\over m}\right)^{2\epsilon}
\left[
{e^{\gamma\epsilon}\Gamma(1+\epsilon)\over\epsilon}
+\left(d-4\right)
J_2+2J_3m^2T^{-2}
\right]
\;,\\
\sumint_{\{P\}}{p_0^2\over(P^2+m^2)^4}&=&{1\over12(4\pi)^2}
\left({\mu\over m}\right)^{2\epsilon}
\left[
e^{\gamma\epsilon}\Gamma(1+\epsilon)m^{-2}
+\left(d-6\right)J_3T^{-2}+2J_4m^2T^{-4}
\right]
\;,\\
\label{p2}
\sumint_{\{P\}}{p_0^2p_ip_j\over(P^2+m^2)^4}&=&{\delta_{ij}\over24(4\pi)^2}
\left({\mu\over m}\right)^{2\epsilon}
\left[
{e^{\gamma\epsilon}\Gamma(1+\epsilon)\over\epsilon}
+\left(d-4\right)
J_2+2J_3m^2T^{-2}
\right]
\;.
\eqa
The integrals $J_n(\beta m)$ 
can be expressed as integrals involving the Fermi-Dirac
distribution function:
\bqa
J_n(\beta m)&=&{4e^{\gamma\epsilon}\Gamma(\mbox{${1\over2})$}
\over\Gamma(\mbox{${5\over2}$}-n-\epsilon)}
\beta^{4-2n}m^{2\epsilon}\int_0^{\infty}
dk\;{k^{4-2n-2\epsilon}\over(k^2+m^2)^{1/2}}{1\over 
e^{\beta(k^2+m^2)^{1/2}}+1}\;.
\eqa
These integrals are functions of $\beta m$ only and satisfy the recursion
relation
\bqa
m{\partial\over\partial m}J_n(\beta m)
=2\epsilon J_n(\beta m)-2(\beta m)^2J_{n+1}(\beta m)\;.
\eqa
We need the integrals $J_n$ for $\epsilon=0$.
In the low-temperature limit, these integrals reduce to
\bqa
J_0&\longrightarrow&8(2\pi)^{1/2}
m^{3/2}T^{-3/2}e^{-m/T}\;,\\
J_1&\longrightarrow&4(2\pi)^{1/2}
m^{1/2}T^{-1/2}e^{-m/T}\;,\\
J_2&\longrightarrow&2(2\pi)^{1/2}
m^{-1/2}T^{1/2}e^{-m/T}\;,\\
J_3&\longrightarrow&(2\pi)^{1/2}
m^{-3/2}T^{3/2}e^{-m/T}\;,\\
J_4&\longrightarrow&{1\over2}(2\pi)^{1/2}
m^{-5/2}T^{5/2}e^{-m/T}\;.
\eqa
Note that the integrals $J_3$ and $J_4$ need subtractions to 
remove power infrared divergences:
\bqa
J_3(\beta m)&=&-2
T^2\int_0^{\infty}
dk\;{1\over k^2}
\left[{1\over(k^2+m^2)^{1/2}}{1\over e^{\beta(k^2+m^2)^{1/2}}+1}
-{1\over m}{1\over e^{\beta(k^2+m^2)^{1/2}}+1}
\right]\;, \\
J_4(\beta m)&=&3\nonumber
T^4\int_0^{\infty}
dk\;{1\over k^4}
\Bigg[{1\over(k^2+m^2)^{1/2}}{1\over e^{\beta(k^2+m^2)^{1/2}}+1}
-{1\over m}{1\over e^{\beta(k^2+m^2)^{1/2}}+1}
\\
&&
\hspace{2.6cm}
+{k^2\over2m^3}{1\over e^{\beta(k^2+m^2)^{1/2}}+1}
\Bigg]\;.
\eqa

The specific one-loop bosonic sum-integral needed is
\bqa
\label{logp}
\sumint_{P}\log(P^2)&=&-{\pi^2T^4\over45}\left[
1+{\cal O}(\epsilon)
\right]\;.
\eqa

\subsection{Two-loop Sum-integral}
We also need the value of the two-loop diagram in Eq.~(\ref{two}).
The two-loop sum-integral with nonzero chemical $\mu$ potential
was calculated in e.g.
Refs.~\cite{kap2,toimela,kapusta}. There are 
$T$-dependent and $\mu$-dependent infinities
in addition to the usual vacuum infinities. These are cancelled by the
corresponding infinities arising from the one-loop counterterm diagrams.
The vacuum infinity is cancelled by a vacuum counterterm in the usual way.
If we demand that the two-loop contribution to the vacuum energy vanishes
at the scale $\Lambda$, the diagram is given by its finite-temperature
piece.
The final result after renormalization is then~\cite{toimela}:
\bqa\nonumber
{1\over2}e^2
\sumint_{\{PQ\}}\mbox{Tr}\left[
\gamma_{\mu}
{P\!\!\!\!/-m\over P^2+m^2}
\gamma_{\mu}{Q\!\!\!\!/-m\over Q^2+m^2}
{1\over(P+Q)^2}
\right]&=&
-{e^2\over12\pi^2}T^2\int_0^{\infty}dp\;{p^2\over E_p}\left[n_p^++n_p^-\right]
\\ \nonumber
&&\hspace{-7cm}
-
{e^2\over16\pi^4}\int_0^{\infty}dp\;dq\;
{p^2q^2\over E_pE_q}
\left[
\left(
2+{m^2\over pq}\log{E_pE_q-m^2-pq\over E_pE_q-m^2+pq}
\right)\left(n_p^-n_q^-+n_p^+n_q^+\right)
\right.
\\
&&\hspace{-7cm}
+
\left.
\left(
2+{m^2\over pq}\log{E_pE_q+m^2+pq\over E_pE_q+m^2-pq}
\right)\left(n_p^-n_q^++n_p^+n_q^-\right)
\right]
\label{fermi2}
\;,
\eqa
where $E_p=\sqrt{p^2+m^2}$ and 
\bqa
n_p^{\pm}={1\over e^{\beta(E_p\pm\mu)}+1}\;.
\eqa
In the low-temperature limit, this reduces to
\bqa
\label{limit}
{1\over2}e^2
\sumint_{\{PQ\}}\mbox{Tr}\left[
\gamma_{\mu}
{P\!\!\!\!/-m\over P^2+m^2}
\gamma_{\mu}{Q\!\!\!\!/-m\over Q^2+m^2}
{1\over(P+Q)^2}
\right]&\longrightarrow&
{e^2\over2(2\pi)^3}m^2T^2e^{2(\mu-m)/T}
\;.
\eqa
Eq.~(\ref{fermi2}) gives the two-loop contribution to the pressure ${\cal P}$.
By applying the formula~\cite{kapusta}
\bqa
\Pi_{00}(0,0)=e^2{\partial^2{\cal P}^2\over\partial\mu^2}\;,
\eqa
to Eq.~(\ref{limit}), we obtain the two-loop contribution to the photon
polarization tensor at zero external momentum and vanishing 
chemical potential:
\bqa
\label{twopol}
\Pi_{00}^{(2)}(0,0)
&=&
{2e^4\over(2\pi)^3}m^2e^{-2m/T}
\;.
\eqa
\section{Integrals}

Dimensional regularization can be used to 
regularize both the ultraviolet divergences and infrared divergences
in 3-dimensional integrals over momenta. 
The spatial dimension is generalized to  $d = 3-2\epsilon$ dimensions.
Integrals are evaluated at a value of $d$ for which they converge and then
analytically continued to $d=3$.
We use the integration measure
\begin{equation}
  \int_{\bf p} \;\equiv\;
  \left(\frac{e^\gamma\mu^2}{4\pi}\right)^\epsilon\,
  \int {d^{3-2\epsilon}p \over (2 \pi)^{3-2\epsilon}}\,,
\end{equation}
where 
$\mu$ is an arbitrary
momentum scale. 
The factor $(e^\gamma/4\pi)^\epsilon$
is introduced so that, after minimal subtraction 
of the poles in $\epsilon$
due to ultraviolet divergences, $\mu$ coincides 
with the renormalization
scale of the $\overline{\rm MS}$ renormalization scheme.

The one-loop integrals required are
\begin{eqnarray}
\int_{\bf p} \log(p^2+m^2) & = & 
- {m^3\over 6\pi}\left[1+{\cal O}(\epsilon)\right]
  \,,
\label{3d}
\\ 
\int_{\bf p} {1 \over p^2+m^2} & = & 
- {m\over 4\pi} 
\left[1 + {\cal O}\left(\epsilon\right)  \right] \,.
\label{bi3}
\end{eqnarray}

\section{Polarization Tensor}
In this appendix, we calculate the polarization tensor
$\Pi_{\mu\nu}(\omega_n,{\bf k})$ to one loop. 
The Feynman diagram is shown in 
Fig.~\ref{pollen} and the expression for it is
\bqa
\label{pol}
\Pi_{\mu\nu}^{(1)}(k_0,{\bf k})=e^2\sumint_{\{P\}}\mbox{Tr}\left[
\gamma_{\mu}{P\!\!\!\!/-m\over P^2+m^2}
\gamma_{\nu}{(P\!\!\!\!/+K\!\!\!\!/)-m\over(P+K)^2+m^2}\right]\;.
\eqa
Taking the Dirac trace, 
and using a Feynman parameter $y$, this can be writen as
\bqa\nonumber
\Pi_{\mu\nu}^{(1)}(k_0,{\bf k})&=&e^2\int_0^1dy\;
\sumint_{\{P\}}
\Bigg[
{8p_{\mu}p_{\nu}\over\left[P^2+m^2+K^2y(1-y)\right]^2}
-{4\delta_{\mu\nu}\over\left[P^2+m^2+K^2y(1-y)\right]}
\\
&&
+{8y(1-y)(\delta_{\mu\nu}K^2-k_{\mu}k_{\nu})
\over\left[P^2+m^2+K^2y(1-y)\right]^2}
\Bigg]\;.
\eqa
First consider $\Pi_{ij}^{(1)}(0,{\bf k})$. By virtue of 
Eqs.~(\ref{spat1}) and~(\ref{spat2}), the first two terms cancel identically.
The remaining term is expanded to second order 
in a Taylor series around ${\bf k}=0$. 
Integrating over $y$, we obtain
\bqa
\Pi_{ij}^{(1)}(0,{\bf k})&=&
{4\over3}
{e^2}(\delta_{ij}k^2-k_{i}k_{j})\sumint_{\{P\}}{1\over (P^2+m^2)^2} 
\label{space}
-{8\over15}{e^2}k^2(\delta_{ij}k^2-k_{i}k_{j})\sumint_{\{P\}}
{1\over (P^2+m^2)^3}\;.
\eqa
The first term gives the one-loop correction to the field normalization
constants, while the second gives the coefficient of the Uehling term.
The fact that $\Pi_{ij}^{(1)}(0,{\bf k})$ vanishes 
in the limit ${\bf k}\rightarrow 0$ reflects the fact that there is no 
screening of static magnetic fields in QED.

Consider next $\Pi_{00}^{(1)}(0,{\bf k})$. 
Expanding to second order in the external momentum ${\bf k}$
and integrating over $y$, yields
\bqa
\label{stat}
\Pi_{00}^{(1)}(0,{\bf k})
&=&4e^2\sumint_{\{P\}}\Bigg[
{2p_0^2\over(P^2+m^2)^2}-{1\over P^2+m^2}\Bigg]
+{1\over3}e^2k^2\sumint_{\{P\}}\Bigg[
{6\over(P^2+m^2)^2}-{8p_0^2\over(P^2+m^2)^3}\Bigg]
\;.
\eqa
The first term gives the one-loop expression for the mass parameter 
(which coincides with the one-loop expression for the Debye mass), while the
second term gives the one-loop correction to the field normalization constant.

\end{document}